\documentclass[a4paper]{ifacmtg}
\usepackage{amsmath} 
\usepackage{amssymb}  
\usepackage[latin1]{inputenc}
\usepackage{graphicx}
\usepackage{array}
\usepackage{dcolumn}
\usepackage{pst-all}
\usepackage{subfigure}

\usepackage{theorem}

\renewcommand{\e}[1]{\cdot 10^{#1}}
\newcommand{\noopsort}[1]{}
\newcommand{\Real}{\mathbb{R}}

\newtheorem{proposition}{Proposition}
\newtheorem{definition}{Definition}

\newtheorem{theorem}{Theorem}

\usepackage{natbib}

\begin{document}
\begin{frontmatter}
\title{A feedback approach to bifurcation analysis in biochemical networks with many parameters}
\author{Steffen Waldherr\thanksref{mail}} \textbf{and}
\author{Frank Allgöwer}
\address{Institute for Systems Theory and Automatic Control\\
University of Stuttgart, Germany \\
\texttt{http://www.ist.uni-stuttgart.de/}}
\thanks[mail]{Corresponding author (\texttt{waldherr@ist.uni-stuttgart.de})}

\begin{abstract}
Feedback circuits in biochemical networks which underly cellular signaling pathways
are important elements in creating complex behavior. A specific aspect thereof is how
stability of equilibrium points depends on model parameters. For biochemical networks,
which are modelled using many parameters, it is typically very difficult to estimate
the influence of parameters on stability. Finding parameters which result in a change in
stability is a key step for a meaningful bifurcation analysis. 
We describe a method based on well known approaches
from control theory, which can locate parameters leading to a
change in stability. The method considers a feedback circuit in the biochemical network
and relates stability properties to the control system obtained by loop--breaking.
The method is applied to a model of a MAPK cascade as an illustrative example.
\end{abstract}
\vspace*{-0.25cm}
\end{frontmatter}

\section{Introduction}
\label{sec:intro}

Feedback circuits are an important structural feature of biochemical networks \citep{TysonOth1978}.
The presence of complex behavior such as
bistability, i.e.\ the existence of several stable equilibria, and sustained oscillations
can be attributed to the presence of feedback circuits \citep{CinquinDem2002a,KaufmanTho2003}.
These types of complex behavior are directly related to how feedback circuits influence stability
properties of equilibria.
In consequence, stability analysis of biochemical networks involving
feedback is a recurring field of interest, and several theoretical results have been
obtained \citep{DibrovZha1982,Thron1991,AngeliSon2004a}.

Models for biochemical networks in cellular signaling typically contain a large number of parameters whose values
are not exactly known and which can even vary due to differential gene expression
(e.g. the concentration of an enzyme) or external influences (e.g. cofactors). 
These parameters often have a considerable influence on stability, which needs
to be evaluated in order to understand the function of a network \citep{EissingWal2007,KimPos2006}.

A classical tool to study the influence of parameter variations on stability is
bifurcation analysis. It has been applied to many cellular signalling systems, such as
the lac operon \citep{YildirimMac2003}
and the MAPK cascade \citep{MarkevichHoe2004,ChickarmaneKho2007}, to name but a few.
When considering models with many parameters, one faces the difficulty that
in classical bifurcation analysis, only one parameter at a time can be varied.
Thus the effect of simultaneous variations in several parameters can not be
evaluated properly.

In this paper, we present a new approach to locate bifurcations in systems with feedback loops
containing many parameters which may be varied simultaneously. To this end, we make use of
an appropriate frequency domain description of the system and of mathematical conditions
representing necessary conditions for a bifurcation.
The paper is structured as follows. In Section~\ref{sec:theory}, we present
the theoretical results required for our approach and suggest an optimization--based method
to actually find interesting parameter values.
In Section~\ref{sec:application}, we apply these results to a
model of the MAPK cascade with a negative feedback circuit \citep{Kholodenko2000}.
The relevance of our results is discussed in Section~\ref{sec:conclusions}.
The mathematical proofs of the theoretical results are not presented in this paper
but will be provided elsewhere (Waldherr and Allgöwer, in preparation).

\section{Theoretical background}
\label{sec:theory}

\subsection{The loop--breaking approach}

We consider a nonlinear differential equation which may describe the biochemical network
constituting a cellular signaling pathway,
\begin{equation}
\label{eq:closed-loop-system}
\begin{aligned}
\Sigma:\ \dot x = F(x,p),
\end{aligned}
\end{equation}
with $x\in\Real^n$ and $p\in\mathcal{P}$, where $\mathcal{P}$ is a connected subset of $\Real^m$.
Typically, $x$ will represent the concentrations
of the signaling molecules, and $p$ collects parameters like reaction constants or enzyme
concentrations.
We assume that an equilibrium point $\bar x(p)$ exists for all parameter values and can be computed at least numerically,
such that $F(\bar x(p),p)=0$ for all $p\in\Real^m$.

Mathematically, the system \eqref{eq:closed-loop-system} is said to contain
a feedback circuit if the influence graph of its Jacobian $\frac{\partial F}{\partial x}$
contains a nontrivial loop \citep{CinquinDem2002a}. We want to study the influence of such a feedback circuit on the
dynamical properties of the system. Control theory provides efficient tools to study this problem.
A useful approach in our setup is to consider
the system \eqref{eq:closed-loop-system} as
a closed loop control system. It is then possible to study the corresponding open loop system, and
one can resort to the rich stability theory developped for control systems. 

An open loop control system corresponding to the closed loop system \eqref{eq:closed-loop-system} is
obtained by loop breaking, as defined in the following.
\begin{definition}
A \emph{loop breaking} for the system \eqref{eq:closed-loop-system} is a pair $(f,h)$, where
$f:\Real^n\times\Real\times\Real^m\rightarrow\Real^n$ is a smooth vector field and $h:\Real^n\rightarrow\Real$
is a smooth function, such that
\begin{equation}
\label{eq:loop-breaking}
\begin{aligned}
F(x,p) = f(x,h(x),p).
\end{aligned}
\end{equation}
\end{definition}

The corresponding open loop system is then given by the equation 
\begin{equation}
\label{eq:open-loop-system}
\sigma:\ \left\lbrace\begin{aligned}
\dot x &= f(x,u,p) \\
y &= h(x).
\end{aligned}\right.
\end{equation}
The closed loop system can again be obtained by ``closing the loop'', i.e.\ setting $u = y$.
Notice that by the assumption that an equilibrium exists for the closed loop system, the
open loop system also has the equilibrium $\bar x(p)$ when choosing $u = h(\bar x(p))$.
This input is denoted as $\bar u(p) = h(\bar x(p))$.

Since our main interest is in stability properties of the equilibrium point
$\bar x(p)$, we can restrict the analysis to the linear approximation of the systems~\eqref{eq:closed-loop-system}
and \eqref{eq:open-loop-system} around the equilibrium point.
By using Laplace transformation, the linear approximation of the open loop system~\eqref{eq:open-loop-system} can
be represented by a linear parameter-dependent transfer function
\begin{equation}
\label{eq:transfer-function}
\begin{aligned}
G(p,s) = \frac{k(p) q(p,s)}{r(p,s)},
\end{aligned}
\end{equation}
where $q$ and $r$ are polynomials in the complex variable $s$ with coefficients
depending on $p$.
As a technical restriction, we assume that the open loop system has no poles or zeros on the imaginary
axis, i.e.\ $r(p,\cdot)$ and $q(p,\cdot)$ are assumed to have no roots on the imaginary axis
for any value of $p\in\mathcal{P}$ throughout this section.

\subsection{Properties of the closed and open loop systems}

Stability of an equilibrium point of the closed loop system
depends on the position of the eigenvalues of the Jacobian $\frac{\partial F}{\partial x}(\bar x(p),p)$.
To characterize these eigenvalues from conditions on the open loop system,
we have the following theorem.
\begin{theorem}
\label{theo:eigenvalues}
Let $A(p) = \frac{\partial f}{\partial x}\left(\bar x(p),\bar u(p),p\right)$
and $A_{cl}(p) = \frac{\partial F}{\partial x}(\bar x(p),p)$.
Assume that $s_0\in\mathbb C$ is not an eigenvalue of $A(p)$. Then $s_0$ is an eigenvalue
of $A_{cl}(p)$, if and only if $G(p,s_0)=1$.
\end{theorem}
The proof of Theorem~\ref{theo:eigenvalues} is based on a representation of $G$
as
\begin{equation*}
\begin{aligned}
G(p,s) = \frac{\det(s I-A_{cl}(p))}{\det(s I-A(p))}+1.
\end{aligned}
\end{equation*}

Parameter values on the border of stability are characterised by the matrix $A_{cl}(p)$ having
eigenvalues on the imaginary axis.
To study the corresponding property in the frequency domain
representation of the open loop system, we introduce the notation of critical frequencies and gains.
\begin{definition}
We say that $\omega_c\in\Real$ is a \emph{critical frequency} and
$k_c\in\Real$ a corresponding \emph{critical gain} for the transfer function $G(p,\cdot)$ \eqref{eq:transfer-function},
if 
\begin{equation}
\label{eq:wcrit-def}
\begin{aligned}
\frac{k_c q(p,j\omega_c)}{r(p,j\omega_c)} = 1.
\end{aligned}
\end{equation}
\end{definition}
In general, different critical frequencies and gains will be obtained for different
values of $p$.

The critical frequencies can be characterized independently of the critical gains.
This result follows from \eqref{eq:wcrit-def}, because the transfer function value at the critical frequency
has to be a real number.
\begin{proposition}
\label{theo:crit-freq}
$\omega_c$ is a critical frequency for $G(p,\cdot)$, if and only if 
\begin{equation}
\label{eq:wcrit-computation}
\begin{aligned}
\operatorname{Im}(q(p,j\omega_c)r(p,-j\omega_c))=0.
\end{aligned}
\end{equation}
There exists a unique corresponding critical gain for any critical frequency $\omega_c$, which is given by
\begin{equation}
\label{eq:kcrit-computation}
\begin{aligned}
k_c(p,\omega_c) = \frac{r(p,j\omega_c)}{q(p,j\omega_c)}.
\end{aligned}
\end{equation}
\end{proposition}

The equation \eqref{eq:wcrit-computation} is a polynomial in $\omega_c$, thus all
critical frequencies can be computed numerically for fixed parameters $p$.
The set of all critical frequencies for the transfer function $G(p,\cdot)$ is given
by
\begin{equation}
\label{eq:critical-freq-set}
\begin{aligned}
\Omega_c(p) = \left\lbrace \omega\in\Real \mid \operatorname{Im}(q(p,j\omega)r(p,-j\omega))=0 \right\rbrace
\end{aligned}
\end{equation}

The concept of critical frequencies and critical gains can be understood intuitively when
considering the Nyquist plot of the transfer function $G(p,j\omega)$. A critical frequency is
any value $\omega$ at which the Nyquist plot crosses the real axis. 
The corresponding critical
gain is the value $k(p) = k_c$ which scales the Nyquist plot in such a way that the crossing point at the
critical frequency is mapped to $1$ in the complex plane.
However, this intuitive way of scaling the Nyquist plot would require to
keep all critical frequencies in $\Omega_c(p)$ constant when varying
parameters, which would be a strong restriction.
The next section presents an approach to overcome this restriction.

\subsection{A minimal set of critical frequencies}

The number of
critical frequencies that exist for a given open loop system is often predefined by the
position of the open loop poles and zeros in the left or right half complex plane.
The following proposition guarantees the existence
of a minimal number of critical frequencies.
\begin{proposition}
\label{prop:minimality}
Let $\alpha = \vert p_+ - p_- + z_- - z_+ \vert$, where $p_+$ ($p_-$) is the number
of poles of $G(p,\cdot)$ in the right (left) half complex plane and $z_+$ ($z_-$) is the number
of zeros of $G(p,\cdot)$ in the right (left) half complex plane.
Then, for any $p\in\mathcal{P}$,
$\Omega_c(p)$ has at least $\alpha$ distinct elements, if $\alpha$ is odd, and at least
$\alpha-1$ distinct elements, if $\alpha$ is even.
\end{proposition}

Since we assumed that $G(p,\cdot)$ has no poles or zeros on the imaginary axis, 
the number $\alpha$ is the same for all parameters $p\in\mathcal{P}$.
Thus, it can be used to characterise a set of critical frequencies as being minimal.
\begin{definition}
Under the assumptions of Prop.~\ref{prop:minimality},
the set of critical frequencies $\Omega_c(p)$ is called \emph{minimal}, if it contains
exactly the minimal number of elements according to Prop.~\ref{prop:minimality}.
\end{definition}

If $\Omega_c(p)$ is minimal, we can label the roots of \eqref{eq:wcrit-computation} in a consistent way,
and write $\Omega_c(p) = \left\lbrace \omega_c^1(p), \omega_c^2(p), \ldots ,\omega_c^\alpha(p) \right\rbrace$,
where the $\omega_c^i$ can be identified with different solution branches of the polynomial equation \eqref{eq:wcrit-computation}.

\subsection{Existence of critical parameter values}

The concept of critical frequencies and gains is now applied to
the problem of how stability depends on parameters. 
We study the problem of finding critical parameters $p_c\in\mathcal{P}$
on the border of stability, i.e.\ such
that the eigenvalues of the Jacobian $A_{cl}(p_c)$
are located on the imaginary axis. 
Then there exist typically parameters
$p_0$ and $p_1$ in a neighborhood of $p_c$ such that the equilibrium $\bar{x}(p_1)$ is
stable and $\bar{x}(p_2)$ is unstable.

The following theorem uses the loop--breaking approach and the concept of
critical frequencies to characterise the existence of critical parameters.
\begin{theorem}
Assume that $\Omega_c(p)$ is minimal for all $p\in\mathcal{P}$. 
Then there exists $p_c\in\mathcal{P}$ such that
$j\omega_c^i(p_c)$ is an eigenvalue of $A_{cl}(p_c)$, if and only
if there exist $p_0,p_1\in\mathcal{P}$ such that $G(p_0,j\omega_c^i(p_0)) \leq 1$
and $G(p_1,j\omega_c^i(p_1)) \geq 1$, for $\omega_c^i(\cdot)\in\Omega_c(\cdot)$
and for any $i\in\lbrace 1,2,\ldots,\alpha\rbrace$.
\end{theorem}

Thus, instead of
having to look at how the $n$ eigenvalues of the closed--loop system change with
parameters, we have reduced this to one number, given by $G(p,j\omega_c^i(p))$,
which contains all information about whether the system changes its stability properties
when changing parameters.
The result is global in the sense that the parameters $p_0$ and $p_1$
can be arbitrarily far apart from each other, still under the given conditions
existence of critical parameters $p_c$ is guaranteed. 

\subsection{Searching for critical parameter values}
\label{ssec:search-critical-param}

The theoretical approach outlined above can be used to search
for parameter values such that the equilibrium point $\bar{x}(p)$ of the
closed loop system~\eqref{eq:closed-loop-system} changes
its stability.
For a biochemical system, there are often
nominal parameters $p_0$, giving rise to the equilibrium point $\bar{x}(p_0)$.
We want to find parameters $p_1$ such that $\bar{x}(p_0)$ and $\bar{x}(p_1)$ have
different stability properties.

In view of the methods presented in this paper, given the open loop transfer function \eqref{eq:transfer-function}, one
first needs to identify the critical frequency that is to be considered.
This choice depends on the type of stability change one is looking for. When taking $\omega_c=0$,
it is possible to search for zero eigenvalues, and if $\omega_c > 0$, nonzero imaginary eigenvalues
may be encountered, typically giving rise to a Hopf bifurcation in the closed loop system.
The nominal transfer function value at the critical frequency is $G(p_0,j\omega_c(p_0))$.
To change stability properties, we will
then define a value $\gamma$ as either $\gamma>1$, if $G(p_0,j\omega_c(p_0)) < 1$, or as $\gamma < 1$
otherwise. Then, any solution to the nonlinear equation
\begin{equation}
\begin{aligned}
G(p_1,j\omega_c(p_1)) = \gamma
\end{aligned}
\end{equation}
gives parameters $p_1$ such that $\bar x(p_0)$ and $\bar x(p_1)$ have different stability
properties as indicated by the chosen critical frequency $\omega_c$. 
This method has been implemented using a nonlinear constrained optimization
algorithm from the \textsc{Matlab} Optimization Toolbox \citep{MathWorks2006}.
It allows to efficiently compute parameter values for the desired transfer function value for medium
sized systems, as the example presented in the following section illustrates.

Once a parameter
$p_1$ is known, we can use a straight line going from $p_0$ to $p_1$, defined as
$p_\mu = p_0 +\mu (p_1 -p_0)$. The change in dynamical behaviour along this line can
then be studied using classical bifurcation analysis with respect to $\mu$, implemented
usually via continuation methods \citep{Kuznetsov1995}. In this study, the software AUTO \citep{DoedelPaf2006}
has been used for the bifurcation analysis along the parameter line $p_\mu$.

\section{Application to a MAPK signaling module}
\label{sec:application}

\subsection{Model description}

The method presented in Section~\ref{sec:theory} has been applied to an ODE model
of a mitogen activated protein kinase (MAPK) signaling module. MAPK signaling is a recurring
motif in cellular signaling pathways, and typically appears in a cascade involving
three levels \citep{PearsonRob2001}.

For this study, we consider a mathematical model for the Ras/Raf signaling pathway
similar to the one presented by \cite{Kholodenko2000}.
The inhibition of the upstream molecule SOS by activated MAPK,
the lowest level in the cascade, constitutes a negative feedback circuit around the cascade.
Via the loop--breaking approach, the influence of this feedback connection on existence
of sustained oscillations in kinase activity is analysed.

\begin{figure}
\begin{center}
\begin{pspicture}(\linewidth,4.5cm)
\rput[bl](0,0){\includegraphics[width=\linewidth]{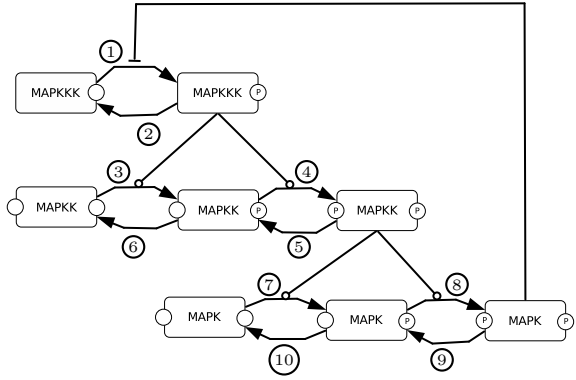}}
\psset{framesep=1pt}
\rput(1.4,3.9){\pscirclebox{\tiny 1}}
\rput(1.9,2.8){\pscirclebox{\tiny 2}}
\rput(1.5,2.3){\pscirclebox{\tiny 3}}
\rput(4.0,2.3){\pscirclebox{\tiny 4}}
\rput(3.9,1.3){\pscirclebox{\tiny 5}}
\rput(1.7,1.3){\pscirclebox{\tiny 6}}
\rput(3.5,0.8){\pscirclebox{\tiny 7}}
\rput(6.0,0.8){\pscirclebox{\tiny 8}}
\rput(5.8,-0.2){\pscirclebox{\tiny 9}}
\rput(3.7,-0.2){\pscirclebox{\tiny 10}}
\end{pspicture}
\end{center}
\vspace*{0.3cm}
\caption{Illustration of the MAPK cascade model.}
\label{fig:mapk-model}
\end{figure}

The structure of the model is illustrated in Fig.~\ref{fig:mapk-model}. The reaction rates
as labeled in the figure are displayed in Table~\ref{tab:mapk-reaction-rates}.
The concentrations have been denoted as $x_{11}=$ [MAPKKK*], $x_{21}=$ [MAPKK*], $x_{22}=$ [MAPKK**],
$x_{31}=$ [MAPK*] and $x_{32}=$ [MAPK**]. The concentrations of unphosphorylated kinases
can be computed by conservation laws and the three parameters $x_{1t}$, $x_{2t}$ and $x_{3t}$
for the total concentrations of the three kinases.
The difference to the model from \cite{Kholodenko2000} is that the phosphorylation reactions
3, 4, 7 and 8 are assumed to follow mass action rather than Michaelis-Menten kinetics. This is
reasonable since the Michaelis-Menten kinetics assumes low enzyme concentration compared to the substrate,
whereas the concentrations of the kinases are in a comparable range here.
Nominal parameter values have been adopted from \cite{Kholodenko2000}, 
and are shown in Table~\ref{tab:mapk-parameters} as $p_0$.

\begin{table}
\renewcommand{\arraystretch}{1.2}
\begin{center}
\begin{tabular}{|c|c|}\firsthline
Reaction & Rate \\ \hline
$v_1$ & $V_1\frac{x_{1t}-x_{11}}{(1+x_{32}/K_{i})(K_{m1}+x_{1t}-x_{11})}$ \\[0.1cm] \hline
$v_2$ & $V_2\frac{x_{11}}{K_{m2}+x_{11}}$ \\[0.1cm] \hline
$v_3$ & $k_3 x_{11}(x_{2t}-x_{21}-x_{22})$ \\[0.1cm] \hline
$v_4$ & $k_4 x_{11} x_{21}$ \\[0.1cm] \hline
$v_5$ & $V_5 \frac{x_{22}}{K_{m5}+x_{22}}$ \\[0.1cm] \hline
$v_6$ & $V_6 \frac{x_{21}}{K_{m6}+x_{21}}$ \\[0.1cm] \hline
$v_7$ & $k_7 x_{22} (x_{3t}-x_{31}-x_{32})$ \\[0.1cm] \hline
$v_8$ & $k_8 x_{22} x_{31}$ \\[0.1cm] \hline
$v_9$ & $V_9 \frac{x_{32}}{K_{m9}+x_{32}}$ \\[0.1cm] \hline
$v_{10}$ & $V_{10}\frac{x_{31}}{K_{m10}+x_{31}}$ \\[0.1cm] \lasthline
\end{tabular} 
\end{center}
\caption{Reaction rates in the MAPK cascade model}
\label{tab:mapk-reaction-rates}
\end{table}

Using the reaction rates from Table~\ref{tab:mapk-reaction-rates}, the model
can be written as a system of five ODEs with 20 parameters:
\begin{equation}
\label{eq:mapk-model}
\begin{aligned}
\dot x_{11} &= v_1 - v_2 \\
\dot x_{21} &= v_3 + v_5 - v_4 - v_6 \\
\dot x_{22} &= v_4 - v_5 \\
\dot x_{31} &= v_7 +  v_9 - v_8 - v_{10} \\
\dot x_{32} &= v_8 - v_9
\end{aligned}
\end{equation}

\begin{figure}
\begin{center}
\includegraphics[width=0.8\linewidth]{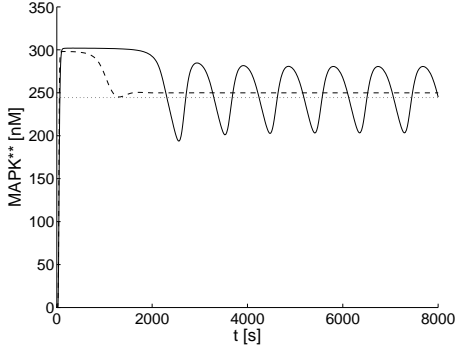}
\end{center}
\caption{Convergence to steady state for parameters $p_0$ (dashed line) and sustained
oscillations for parameters $p_1$ (solid line). The oscillations coexist with an unstable
equilibrium (dotted line).}
\label{fig:mapk-solutions-p0}
\end{figure}

For the nominal parameters $p_0$, the model has a stable equilibrium $\bar x(p_0)$.
Solutions of the model converge quickly to the steady state, as depicted in Fig.~\ref{fig:mapk-solutions-p0}.

\subsection{Parameters for a change in stability properties}

This section describes the application of the method presented in
Section~\ref{sec:theory} to the problem of finding destabilizing parameters
for the MAPK cascade model~\eqref{eq:mapk-model}.

The first step is to choose a suitable loop breaking. For the MAPK cascade, an intuitive
approach is to break the loop at the feedback inhibition of reaction $v_1$ by MAPK**.
Thus we choose $h(x) = x_{32}$, to select [MAPK**] as an output, and replace $x_{32}$
by the input $u$ in the reaction rate $v_1$ to obtain the dynamics of the open loop system $f(x,u,p)$.

\begin{figure}
\begin{center}
\begin{pspicture}(8,2.6)
\rput(2,1.3){\includegraphics[width=0.48\linewidth]{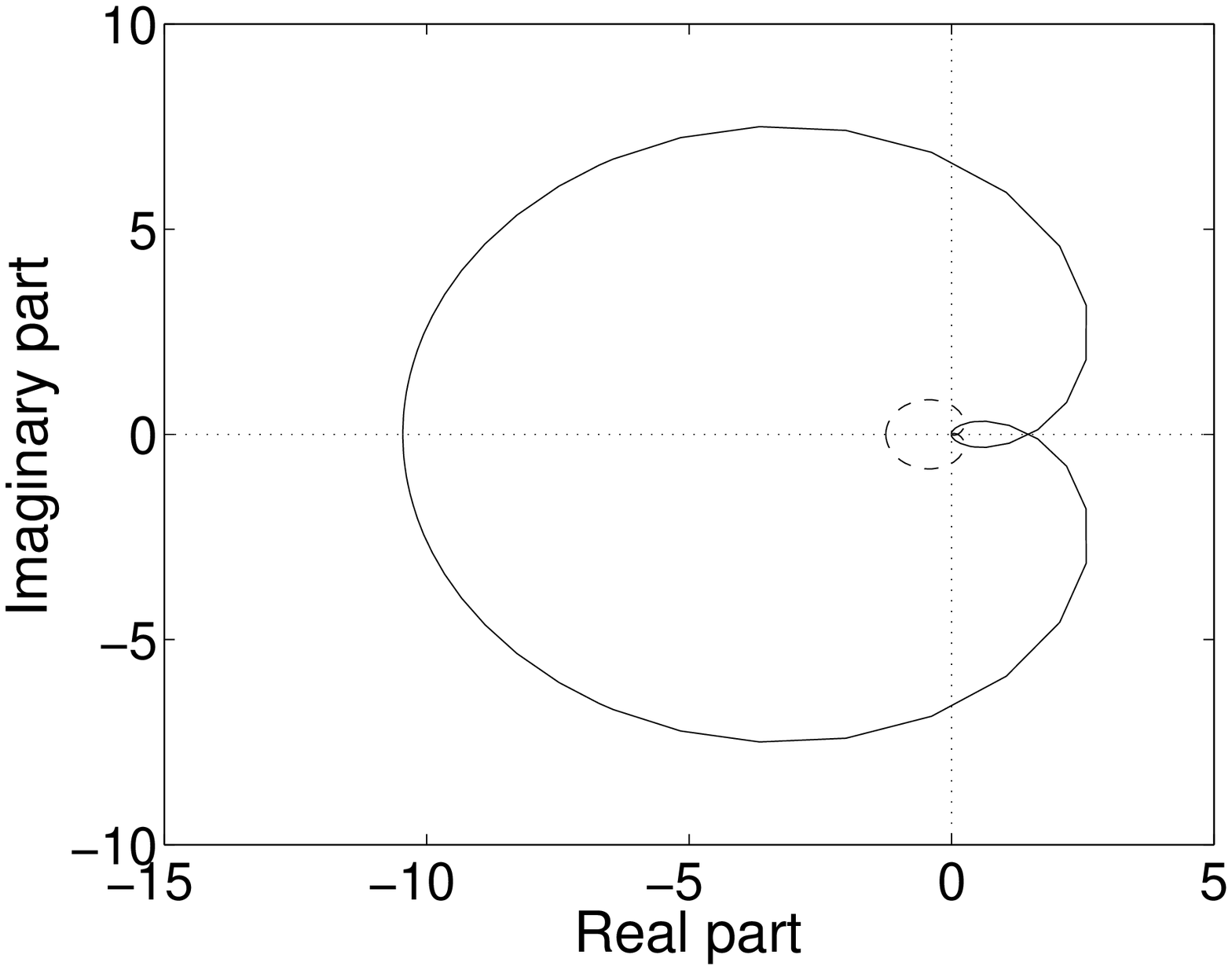}}
\rput(6,1.3){\includegraphics[width=0.48\linewidth]{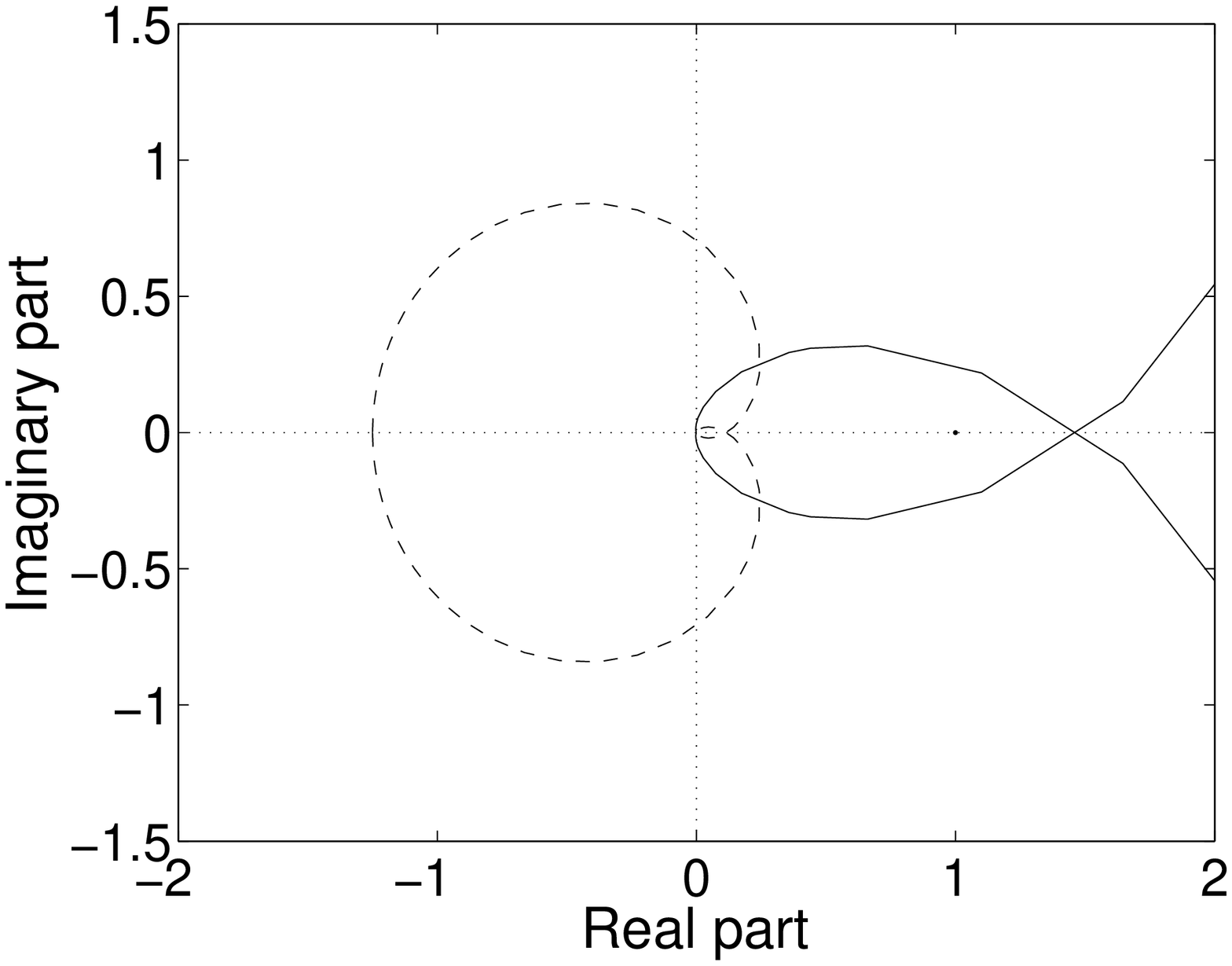}}
\psset{linewidth=0.006}
\psframe(2.65,1.2)(3.3,1.62)
\psset{linewidth=0.001,linestyle=dashed}
\psline(2.65,1.2)(4.72,0.24)
\psline(2.65,1.62)(4.71,2.6)
\psline(3.3,1.2)(7.73,0.24)
\psline(3.3,1.62)(7.73,2.6)
\end{pspicture}
\end{center}
\caption{Nyquist plots of open--loop MAPK model for parameters $p_0$ (dashed line) and $p_1$ (solid line).}
\label{fig:mapk-nyquist}
\end{figure}

It can be shown that there is a unique equilibrium of \eqref{eq:mapk-model} for any parameters in the biologically
meaningful range. The equilibrium can easily be computed numerically. A linearisation of the open loop system
around this equilibrium point and a Laplace transformation gives the transfer function 
$G(p,s)$, whose graph is shown in Fig.~\ref{fig:mapk-nyquist}.
The set of critical frequencies is minimal with $\alpha=3$, which can be seen from Fig.~\ref{fig:mapk-nyquist} by
the observation that the graph of $G(p_0,j\omega)$ encircles the origin monotonically.
The only positive critical frequency is $\omega_c(p_0) = 0.017 s^{-1}$, and we will
consider this frequency in the search for destabilizing parameters. The corresponding transfer function value
is $G(p_0,j\omega_c(p_0)) = 0.12$, corresponding to the equilibrium $\bar x(p_0)$ being stable
in the closed loop system.

For the computational approach described in Section~\ref{ssec:search-critical-param}, we chose
$\gamma = 1.5$, such that the value of the transfer function would have to pass the point 1 when
going from its inital value of $0.12$ to $\gamma$. The optimization method converges to the parameters
$p_1$, which give the desired value $G(p_1,j\omega_c(p_1)) = 1.5$ at a critical frequency
$\omega_c(p_1) = 0.0065 s^{-1}$. The parameter values in $p_1$ are shown in Table~\ref{tab:mapk-parameters}.
The maximal single parameter change from $p_0$ to $p_1$ has been restricted in the numerical implementation
to be not more than a factor of $5$. Even with this restriction, parameters leading to sustained oscillations
have been found.
However, 11 out of the 20 parameters have been changed by more than 20 \% to achieve this.

\begin{table}
\begin{center}
\begin{tabular}{|c|c|c|c|l|} \hline
Param. & $p_0$ & $p_1$ & Unit & rel. change\\ \hline
   $V_1$ & 2.5 & 2.4 & nM/s & $1.05^{-1}$\\ \hline
   $K_i$ & 9 & 10.6 & nM & $1.18$\\ \hline
  $K_{m1}$ & 10 & 9.4 & nM & $1.06^{-1}$\\ \hline
   $V_{2}$ & 0.25 & 0.11 & nM/s & $2.24^{-1}$\\ \hline
  $K_{m2}$ & 8 & 1.6 & nM & $4.9^{-1}$\\ \hline
   $k_{3}$ & 0.001 & 0.0026 & 1/(s nM) & $2.6$\\ \hline
   $k_{4}$ &  0.001 & $3.5\e{-4}$ & 1/(s nM) & $2.8^{-1}$\\ \hline
   $V_{5}$ & 0.75 & 0.32 & nM/s & $2.35^{-1}$\\ \hline
  $K_{m5}$ & 15 & 3.9 & nM & $3.8^{-1}$\\ \hline
   $V_{6}$ & 0.75 & 3.7 & nM/s & $5.0$\\ \hline
  $K_{m6}$ & 15 & 13.3 & nM & $1.12^{-1}$\\ \hline
   $k_{7}$ & 0.001 & 0.0033 & 1/(s nM) & $3.3$\\ \hline
   $k_{8}$ & 0.001 & $5.0\e{-4}$ & 1/(s nM) & $2.00^{-1}$\\ \hline
   $V_{9}$ & 0.5 & 0.26 & nM/s & $1.92^{-1}$\\ \hline
  $K_{m9}$ & 15 & 14.9 & nM & $1.01^{-1}$\\ \hline
  $V_{10}$ & 0.5 & 2.5 & nM/s & $5.0$\\ \hline
 $K_{m10}$ & 15 & 15.0 & nM & $1.00$\\ \hline
  $x_{1t}$ & 100 & 100.0 & nM & $1.00$\\ \hline
  $x_{2t}$ & 300 & 300.8 & nM & $1.00$\\ \hline
  $x_{3t}$ & 300 & 304.2 & nM & $1.01$\\ \hline
\end{tabular} 
\end{center}
\caption{Reference parameters $p_0$ and parameters for instability $p_1$ in the
MAPK cascade model.}
\label{tab:mapk-parameters}
\end{table}

The graph of $G(p_1,j\omega)$ is shown in Figure~\ref{fig:mapk-nyquist}. For the new parameters $p_1$,
the graph now encircles the point 1. By the argument principle, we see
that the linearisation of the closed--loop system around the equilibrium has some eigenvalues
in the right half complex plane and is thus unstable. The sustained oscillations that
appear in this case are shown in Fig.~\ref{fig:mapk-solutions-p0}.

In conclusion, our method is able to compute parameters which render the stable equilibrium unstable
and thus lead to the emergence of sustained oscillations. About half of the parameters are
varied by a non--negligible amount, but all variations are within the physiological range.

\subsection{Bifurcation analysis along a line}

\begin{figure}
\begin{center}
\includegraphics[width=0.75\linewidth]{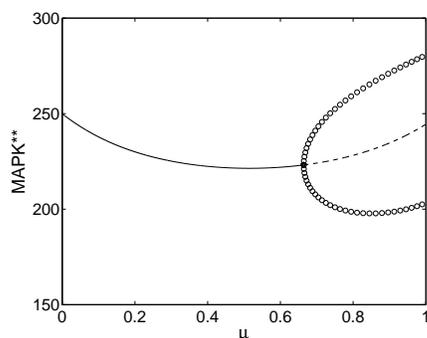}
\end{center}
\caption{Bifurcation diagram along the line $p_\mu$, showing stable equilibrium (solid line),
unstable equilibrium (dashed line) and amplitude of oscillations (circles).}
\label{fig:mapk-bif-diagram}
\end{figure}

Let us now consider
the line $p_\mu = p_0 + \mu (p_1 - p_0)$. By classical bifurcation analysis with $\mu$ as
bifurcation parameter, we can see how the system changes from the stable to the unstable
equilibrium. The resulting bifurcation diagram is shown in Figure~\ref{fig:mapk-bif-diagram}.
As expected, there is a Hopf bifurcation between $p_0$ and $p_1$, at $\mu = 0.664$.
The evolution of the limit cycle producing the sustained oscillations along the line
in parameter space is obtained from the bifurcation diagram.

\section{Conclusions}
\label{sec:conclusions}

We introduced some theoretical tools to investigate the existence of parameters for which a
bifurcation can occur in a dynamical system with a feedback circuit. These tools gave rise
to a new computational method which allows to search for parameter values such that the stability
properties of an equilibrium change in a specific way compared to the nominal parameter values.
Our approach is particularly useful if there are many parameters in the system which can
be varied simultaneously, and if the contribution of individual parameters to stability
properties is not obvious. The ability to directly
handle multiparametric variations is a clear advantage compared to using only classical bifurcation analysis.

We have shown the application of the proposed method to a model of a MAPK cascade.
Using relatively small changes to most of the 20 parameters in the model leads to
a change from a stable equilibrium to an unstable equilibrium with a stable limit cycle,
producing sustained oscillations.

\begin{small}
\bibliography{/home/waldherr/Forschung/Referenzen}

\begin{thebibliography}{16}
\providecommand{\natexlab}[1]{#1}
\providecommand{\url}[1]{\texttt{#1}}
\expandafter\ifx\csname urlstyle\endcsname\relax
  \providecommand{\doi}[1]{doi: #1}\else
  \providecommand{\doi}{doi: \begingroup \urlstyle{rm}\Url}\fi

\bibitem[Angeli and Sontag(2004)]{AngeliSon2004a}
D.~Angeli and E.~D. Sontag.
\newblock Interconnections of monotone systems with steady-state
  characteristics.
\newblock In M.~de~Queiroz, M.~Malisoff, and P.~Wolenski, editors,
  \emph{Optimal control, stabilization, and nonsmooth analysis}, pages
  135--154. Springer-Verlag, 2004.

\bibitem[Chickarmane et~al.(2007)Chickarmane, Kholodenko, and
  Sauro]{ChickarmaneKho2007}
V.~Chickarmane, B.~N. Kholodenko, and H.~M. Sauro.
\newblock Oscillatory dynamics arising from competitive inhibition and
  multisite phosphorylation.
\newblock \emph{J. Theor. Biol.}, 244\penalty0 (1):\penalty0 68--76, January
  2007.
\newblock URL \url{http://dx.doi.org/10.1016/j.jtbi.2006.05.013}.

\bibitem[Cinquin and Demongeot(2002)]{CinquinDem2002a}
O.~Cinquin and J.~Demongeot.
\newblock Positive and negative feedback: striking a balance between necessary
  antagonists.
\newblock \emph{J. Theor. Biol.}, 216\penalty0 (2):\penalty0 229--241, May
  2002.
\newblock \doi{10.1006/jtbi.2002.2544}.
\newblock URL \url{http://dx.doi.org/10.1006/jtbi.2002.2544}.

\bibitem[Dibrov et~al.(1982)Dibrov, Zhabotinsky, and Kholodenko]{DibrovZha1982}
B.~F. Dibrov, A.~M. Zhabotinsky, and B.~N. Kholodenko.
\newblock {D}ynamic stability of steady states and static stabilization in
  unbranched metabolic pathways.
\newblock \emph{J. Math. Biol.}, 15\penalty0 (1):\penalty0 51--63, 1982.

\bibitem[Doedel et~al.(2006)Doedel, Paffenroth, Champneys, Fairgrieve,
  Kuznetsov, Oldeman, Sandstede, and Wang]{DoedelPaf2006}
E.~J. Doedel, R.~C. Paffenroth, A.~R. Champneys, T.~F. Fairgrieve, Y.~A.
  Kuznetsov, B.~E. Oldeman, B.~Sandstede, and X.~Wang.
\newblock \emph{{AUTO} 2000: continuation and bifurcation software for ordinary
  differential equations}.
\newblock Concordia University, Montreal, Canada, 2006.

\bibitem[Ei{\ss}ing et~al.(2007)Ei{\ss}ing, Waldherr, Allg{\"o}wer, Scheurich,
  and Bullinger]{EissingWal2007}
T.~Ei{\ss}ing, S.~Waldherr, F.~Allg{\"o}wer, P.~Scheurich, and E.~Bullinger.
\newblock {S}teady state and (bi-) stability evaluation of simple protease
  signalling networks.
\newblock \emph{BioSystems}, Epub ahead of print, 2007.
\newblock URL \url{http://dx.doi.org/10.1016/j.biosystems.2007.01.003}.

\bibitem[Kaufman and Thomas(2003)]{KaufmanTho2003}
M.~Kaufman and R.~Thomas.
\newblock Emergence of complex behaviour from simple circuit structures.
\newblock \emph{Comptes rend. biol.}, 326:\penalty0 205--214, 2003.

\bibitem[Kholodenko(2000)]{Kholodenko2000}
B.~N. Kholodenko.
\newblock Negative feedback and ultrasensitivity can bring about oscillations
  in the mitogen-activated protein kinase cascades.
\newblock \emph{Eur. J. Biochem.}, 267\penalty0 (6):\penalty0 1583--88, Mar
  2000.

\bibitem[Kim et~al.(2006)Kim, Bates, Postlethwaite, Ma, and
  Iglesias]{KimPos2006}
J.~Kim, D.~G. Bates, I.~Postlethwaite, L.~Ma, and P.~A. Iglesias.
\newblock Robustness analysis of biochemical network models.
\newblock \emph{IEE Proc. Syst. Biol.}, 153\penalty0 (3):\penalty0 96--104, May
  2006.

\bibitem[Kuznetsov(1995)]{Kuznetsov1995}
Y.~A. Kuznetsov.
\newblock \emph{Elements of Applied Bifurcation Theory}.
\newblock Springer-Verlag New York, 1995.

\bibitem[Markevich et~al.(2004)Markevich, Hoek, and
  Kholodenko]{MarkevichHoe2004}
N.~I. Markevich, J.~B. Hoek, and B.~N. Kholodenko.
\newblock {S}ignaling switches and bistability arising from multisite
  phosphorylation in protein kinase cascades.
\newblock \emph{J. Cell Biol.}, 164\penalty0 (3):\penalty0 353--359, Feb 2004.
\newblock \doi{10.1083/jcb.200308060}.
\newblock URL \url{http://dx.doi.org/10.1083/jcb.200308060}.

\bibitem[{\noopsort{MathWorks}The MathWorks Inc.}(2006)]{MathWorks2006}
\emph{Optimization toolbox. For use with \textsc{Matlab}}.
\newblock {\noopsort{MathWorks}The MathWorks Inc.}, 2006.

\bibitem[Pearson et~al.(2001)Pearson, Robinson, Gibson, Xu, Karandikar, Berman,
  and Cobb]{PearsonRob2001}
G.~Pearson, F.~Robinson, T.~B. Gibson, B.~E. Xu, M.~Karandikar, K.~Berman, and
  M.~H. Cobb.
\newblock {M}itogen-activated protein ({MAP}) kinase pathways: regulation and
  physiological functions.
\newblock \emph{Endocr. Rev.}, 22\penalty0 (2):\penalty0 153--183, Apr 2001.

\bibitem[Thron(1991)]{Thron1991}
C.~D. Thron.
\newblock The secant condition for instability in biochemical feedback-control.
  1. {T}he role of cooperativity and saturability.
\newblock \emph{Bull.\ Math.\ Biol.}, 53\penalty0 (3):\penalty0 383--401, 1991.

\bibitem[Tyson and Othmer(1978)]{TysonOth1978}
J.~J. Tyson and H.~G. Othmer.
\newblock The dynamics of feedback control circuits in biochemical pathways.
\newblock \emph{Progr. Theor. Biol.}, 5:\penalty0 2--62, 1978.

\bibitem[Yildirim and Mackey(2003)]{YildirimMac2003}
N.~Yildirim and M.~C. Mackey.
\newblock Feedback regulation in the lactose operon: a mathematical modeling
  study and comparison with experimental data.
\newblock \emph{Biophys. J.}, 84\penalty0 (5):\penalty0 2841--51, May 2003.

\end{thebibliography}
\end{small}

\end{document}